\documentclass[runningheads]{lncs}
\usepackage[english]{babel}
\usepackage[T1]{fontenc}
\usepackage{carlito}  
\usepackage{newtxtext,newtxmath}  
\usepackage{graphicx}
\usepackage{setspace}
\usepackage{cite}
\usepackage{amsmath}
\usepackage{amsfonts}
\usepackage{newtxtext,newtxmath}
\usepackage[linesnumbered,ruled,vlined]{algorithm2e}
\usepackage{caption}
\usepackage{subcaption}
\usepackage{hyperref}
\usepackage{booktabs}
\usepackage{multirow}
\usepackage{enumitem}
\usepackage{adjustbox}
\usepackage{array}
\usepackage{lipsum}
\usepackage{color}
\usepackage{makecell}

\begin{document}

\title{Bounded confidence dynamics generates opinion cascades on growing scale-free networks}
\titlerunning{Bounded confidence dynamics generates opinioncascades on a growing scale-free network}
%
\author{David Hernandez\inst{1}\orcidID{0000-1111-2222-3333} \and Guillaume Deffuant\inst{2,3}\orcidID{1111-2222-3333-4444} \and Yerali Gandica\inst{3}\orcidID{2222--3333-4444-5555}}

\author{David Hernandez\inst{1} \and
Guillaume Deffuant\inst{2} \and
Yerali Gandica\inst{1,3}}

\authorrunning{D. Hernandez et al.}

\institute{Valencian International University (VIU), Valencia, Spain \and Université Clermont-Auvergne, INRAE, UR LISC, France
 \and
Faculty of Economic and Business Sciences, Department of Economic Analysis: Quantitative Economics, Universidad Autónoma de Madrid, 28049 Madrid, Spain}

\maketitle              
\begin{abstract}
We study the pairwise bounded confidence model on scale-free networks where new agents regularly arrive over time. The probability that arriving agents 
form links to preexisting ones depends on both agent degree and opinion proximity. 
In parameter value ranges where both factors impact the link choice, a new  phenomenon is observed. Minor clusters continuously form on the periphery of the opinion space and remain stable for a time, before suddenly merging with a major cluster in the network. 
We label these processes as "opinion cascades", and analyse their origin and behavior. They are triggered by the arrival of agents acting as "bridges" between the previously disconnected minor and major clusters. 
Lastly, we propose theoretical approximations to describe the varying shapes and merging behavior of opinion cascades under different conditions. 
\end{abstract}
%

\section{Introduction}
\label{intro}

Information cascades are present in a wide range of contexts, such as fads and conventions, (political) decision making processes, opinion dynamics, and even media sharing on social media \cite{Bikhchandani1992}. 
They usually indicate a massive and rather sudden adoption of a new behavior, opinion or political position within a group of individuals. Cascades can carry negative consequences, such as the spreading of misinformation or fake news, which can be fatal in delicate socioeconomic contexts and political struggles \cite{DelVicario2016,Rabb2022}. A better understanding of information cascades is crucial to understand their short and long-term effects.

Modelling information cascades as epidemic spreadings is a common method.
In these models, agents can have different states, such as infected, susceptible, and immune (to the epidemic in question). 
The state of a given agent could change according to stochastic rules involving its neighboring agents 
(see e.g. \cite{Watts2002,Newman2018}). Epidemic spreading models have been extensively analysed with various network types \cite{Centola2007,Gandica2010}, yielding satisfactory results about disease propagation data \cite{Karsai2011,Choloniewski2019,Gleeson2016,Jin2021,Rabb2022}. 

Although some epidemic propagation models include opinion dynamics \cite{Rabb2022}, to the best of our knowledge, cascading phenomena have not been observed when using bounded confidence models in opinion dynamics yet. Research on these models typically focuses on the processes leading to the consensus, polarisation, or multiplicity of opinions. 

In this paper, we use the pairwise bounded confidence model \cite{Deffuant2000,Hegselmann2002}, which takes inspiration from established results in social psychology \cite{Hovland1980}. It proposes one-to-one interactions between agents with relatively close opinions. Upon interacting, both of their opinions get closer to each other. 

When all the agents hold the same confidence bound, pairwise interactions on bounded confidence models converge to one or more separate clusters of opinions, depending on the value of the confidence bound $\epsilon$. On fully-connected networks starting with opinions uniformly drawn in $[0,1]$, the number of resulting clusters is approximately the integer part of $1/2 \epsilon$. On other network types, the consensus happens when $\epsilon \geq 0.5$ \cite{FORTUNATO2004}, but the number of resulting clusters can vary when $\epsilon<0.5$. On scale-free networks, hub nodes take the opinion values of the main $1/2 \epsilon$ clusters \cite{STAUFFER2004}.

Generally in these models, the agent population and the network are fixed. On the contrary, in \cite{GANDICA2024} we studied the pairwise bounded confidence model on a growing fully connected population. Aside from the expected opinion clusters, we observed that secondary clusters appear and remain stable over time as new agents arrive. The newly formed clusters are located approximately at an $\epsilon$ (confidence bound) distance from the main clusters. By contrast, when the population is fixed, secondary clusters are not always present, and when they appear, their size is generally non-significant, explaining why they are often overlooked.

We now analyse the pairwise bounded confidence model on a growing scale-free network. Moreover, homophily is present on the network topology, meaning that links between nodes are determined by a trade-off between node degree and opinion proximity, as in \cite{Gargiulo2017}, and we observe a new phenomenon in the network. 

In parameter value ranges where both the degree and opinion proximity impact the link choice, minor clusters continuously form on the periphery of the opinion space and remain stable for a time, before suddenly merging with a major cluster in the network. This merging process is relatively sudden, and can be interpreted as a cascade of opinions. We show that it is triggered by the arrival of nodes that act as ``bridges" in the network, connecting the previously disconnected clusters and starting the abrupt merging phase. 

This process is also studied through specific opinion dynamics simulations, controlling the initial positions and sizes of the main and secondary clusters. In this way, we compare simulation results on arrival of bridges agents with an approximate theoretical model. 

The following section presents the model and describes it in detail. Afterward, we show the presence of various cascading patterns and community structures in simulation results. Then, we propose a theoretical explanation of the cascades and validate it with specific experiments. The last section of the paper is focused on result discussion and concluding remarks.

\section{Model}

The model includes two processes taking place simultaneously. First, agents can interact and influence each other's opinion. Second, new agents periodically arrive to the network and form links to preexisting agents. 

Initially, the network has $N_0$ fully connected agents. Each agent $i$ holds an opinion $\theta_i(t) \in [0,1]$, which will change upon interacting with another agent. Each interaction consists of choosing at random a pair of connected agents that influence each other's opinion according to the Bounded Confidence model, as described in algorithm \ref{Algo:model} and in \cite{Deffuant2000}. All agents hold the same confidence bound $\epsilon$.

After an interaction, there is a probability $\frac{\delta }{N}$ that a new agent is added to the network ($\delta \leq N_0$ is a parameter and $N$ is the current network size). New agents form links to $m$ preexisting nodes. Function $\varphi_N(i)$ describes the probability of connecting an existing agent $i$ to an arriving $N$th agent:

\begin{align}
    \varphi_N(i) = \frac{ k_i \exp\left(-\beta \lvert \theta_N - \theta_i \rvert\right)} {\sum_{j=1}^{N-1} k_j \exp\left(-\beta \lvert \theta_N - \theta_j \rvert\right)},
\end{align}

where $k_i$ is the degree of the preexisting agent $i$, $\theta_i$ is its opinion, $\theta_N$ is the $N$th agent's opinion, and $\beta$ is the parameter tuning the homophily effect, as defined in \cite{Gargiulo2017}. 

In this way, both the degree of the pre-existing node $i$ and the proximity between opinions have an impact on the probability of forming a link, therefore introducing homophily into the network topology. 

Note that if $\beta = 0$, the network follows the classical preferential attachment in Barabasi-Albert networks, whereas if $\beta \to \infty$, the degree becomes irrelevant and only the opinion difference matters. We focus on intermediate values of parameter $\beta$ for which different trade-offs between opinion proximity and degree determine the network.

Algorithm \ref{Algo:model} provides the details of the model. Note that one interaction advances $t$ by $1/N$ ($N$ being the network size). In this way, as the network grows in size, more interactions take place during a given period of time, i.e, each agent always interacts once on average during one unit of time (i.e. between $t$ and $t+1$).

\begin{algorithm}[h]
\SetKwFunction{createNetwork}{createNetwork}
\SetKwFunction{randomConnectedPair}{randomConnectedPair}
\SetKwFunction{random}{random}
\SetKwFunction{addLinks}{addLinks}
\KwIn{$N_0 \in \mathbb{N}$ initial number of agents, $T \in \mathbb{N}$ time limit, $\delta \in \mathbb{R}^*_+$ average number of agents added between $t$ and $t+1$, $\mu = 0.5$ interaction-tuning parameter.}	
	$ \mathcal{N} \leftarrow \createNetwork(N_0)$\tcp*[f]{Create network with $N_0$ agents}\\
	$N \leftarrow N_0$\\
    \While{$t < T$}{
    $(i,j) \leftarrow \randomConnectedPair(\mathcal{N})$\\
    \If{$|\theta_i - \theta_j | < \epsilon$} 
    {$
    \theta_i \leftarrow \theta_i + \mu(\theta_j - \theta_i)$\tcp*[f]{Bounded confidence interaction}\\
    $\theta_j \leftarrow \theta_j + \mu(\theta_i - \theta_j)$ \\
    }
    \If{$\random(0,1) < 1 / N$}
    {$\theta_{N+1} \leftarrow \random(0,1)$ \tcp*[f]{Add new agent with random opinion}\\
    \For{$1 \leq i \leq N$} 
    {$\varphi_i \leftarrow  k_i \exp\left(-\beta \lvert \theta_N - \theta_i \rvert\right)/ \left( \sum_{j=1}^{N-1} k_j \exp\left(-\beta \lvert \theta_N - \theta_j \rvert\right) \right) $ }
    \addLinks$(m,(\varphi_1,...,\varphi_N), \mathcal{N})$ \tcp*[f]{Add $m$ links with probability $\varphi_i$}\\
    $N \leftarrow N+1$}
    $t \leftarrow t+ 1/N$
    }
\caption{Pseudo-code of the model.}
\label{Algo:model}
\end{algorithm}

Furthermore, during a time unit, $\delta$ new agents arrive into the network on average. Since time increases by $\frac{1}{N}$ after each interaction and there is a probability $\delta/N$ to add a new node, a constant number $\delta$ of new agents arrives to the network between $t$ and $t+1$. This can be related to the approach in Barabasi-Albert networks, where the number of incoming nodes is constant and not influenced by the network size.

\section{Simulation results}

\subsection{Minor clusters merging with major clusters in opinion cascades}

\begin{figure}[h]
    \centering
    \begin{tabular}{m{0.3cm}cc}
    & (a) $\beta = 15$ & (b) $\beta = 20$\\
        \rotatebox{90}{\hspace{0.0 cm} Opinion} & \makecell{\includegraphics[width=0.45\linewidth]{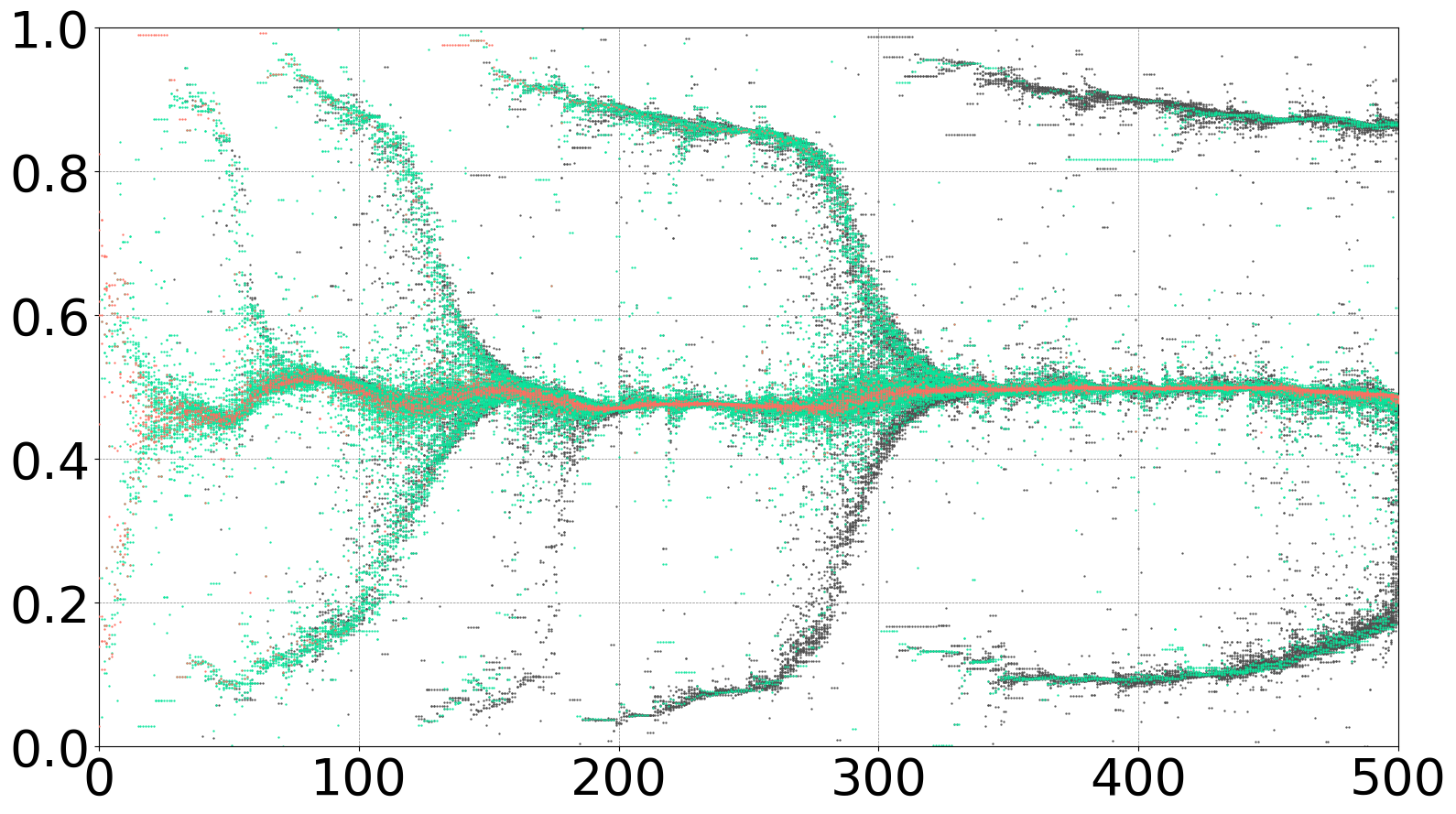}} & 
        \makecell{\includegraphics[width=0.45\linewidth]{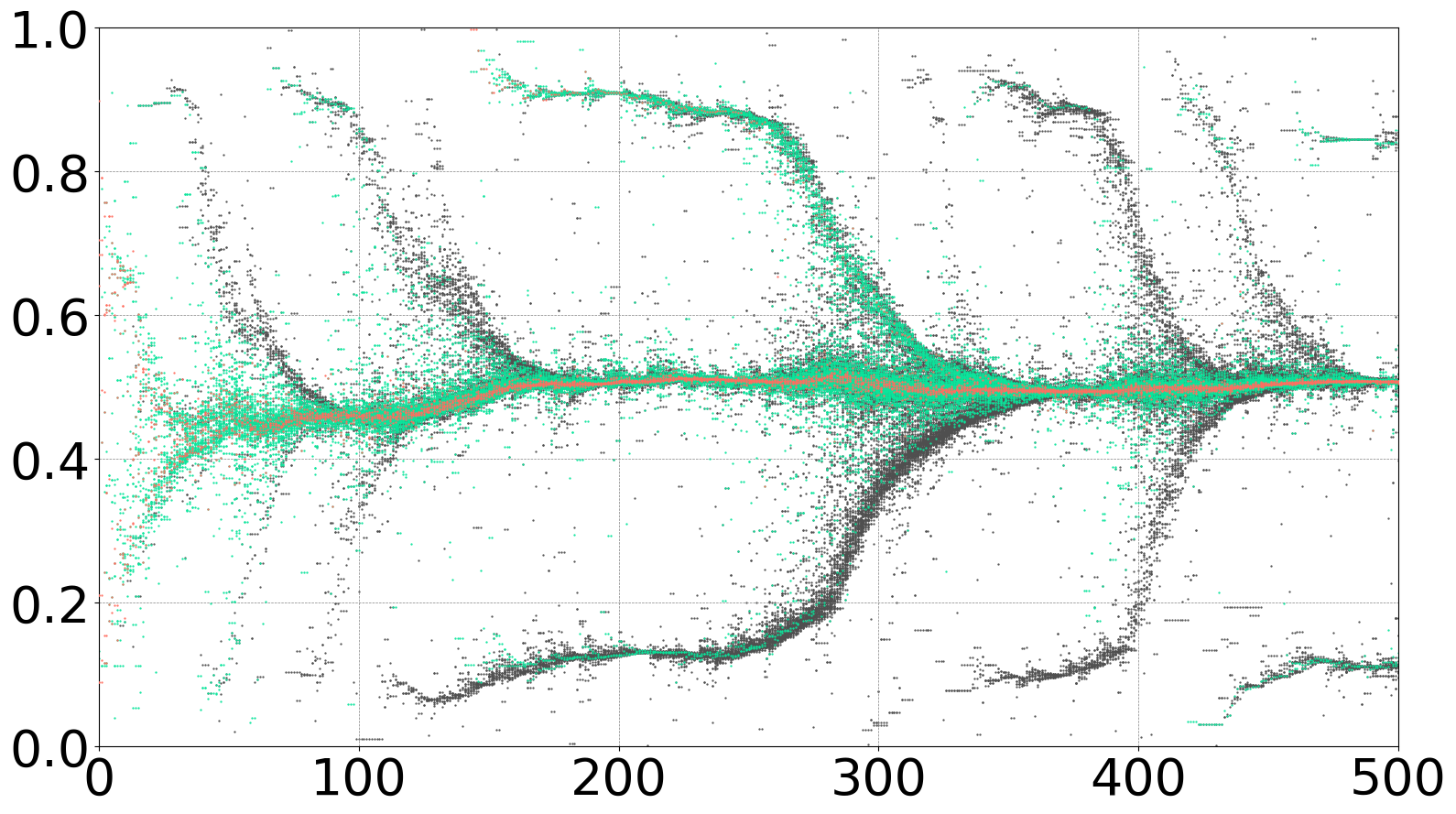}} \\
        & Time & Time \\
        &&\\
         & (c) $\beta = 25$ & (d) $\beta = 30$\\
        \rotatebox{90}{\hspace{0.0 cm} Opinion} & \makecell{\includegraphics[width=0.45\linewidth]{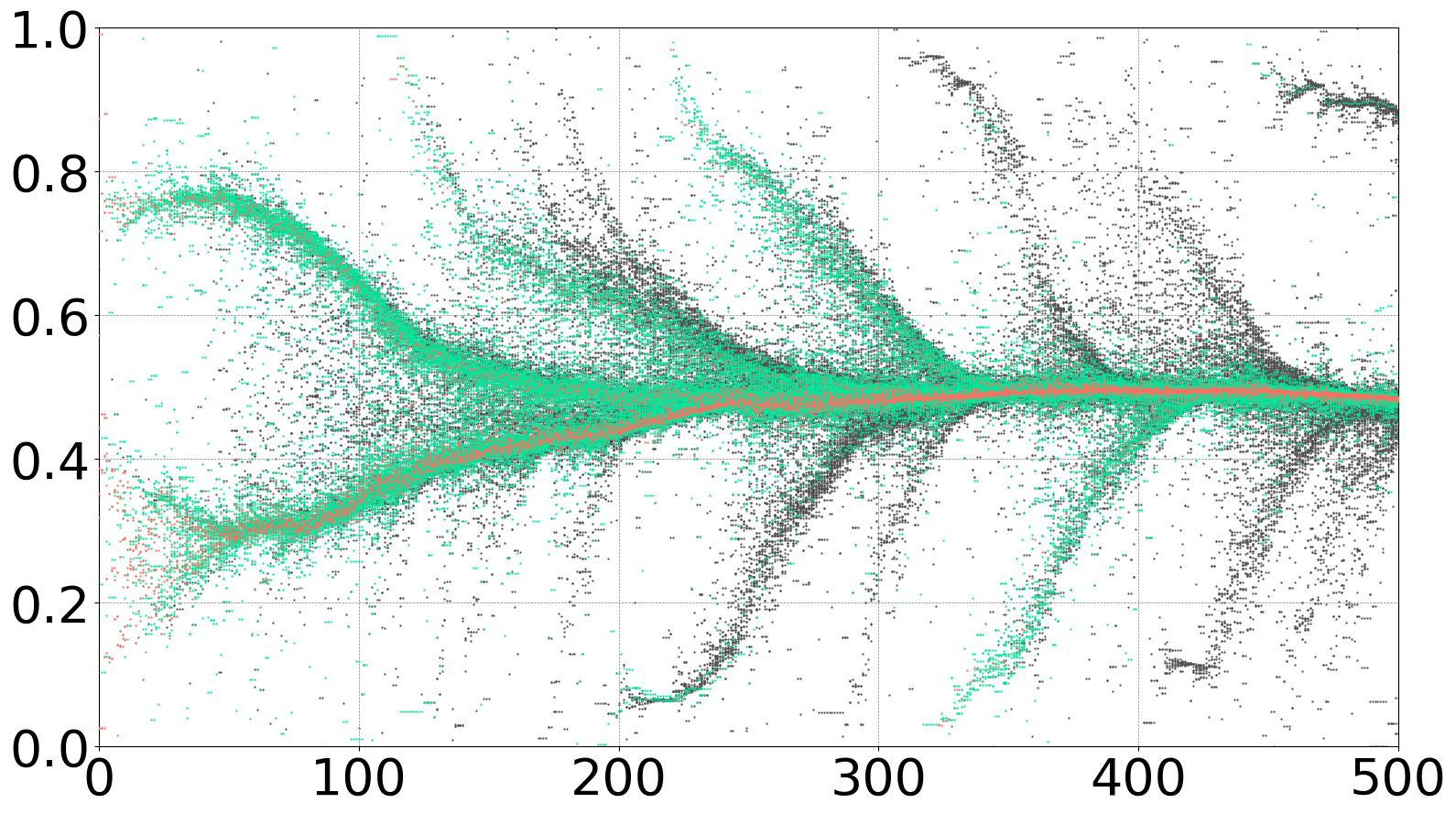}} & 
        \makecell{\includegraphics[width=0.45\linewidth]{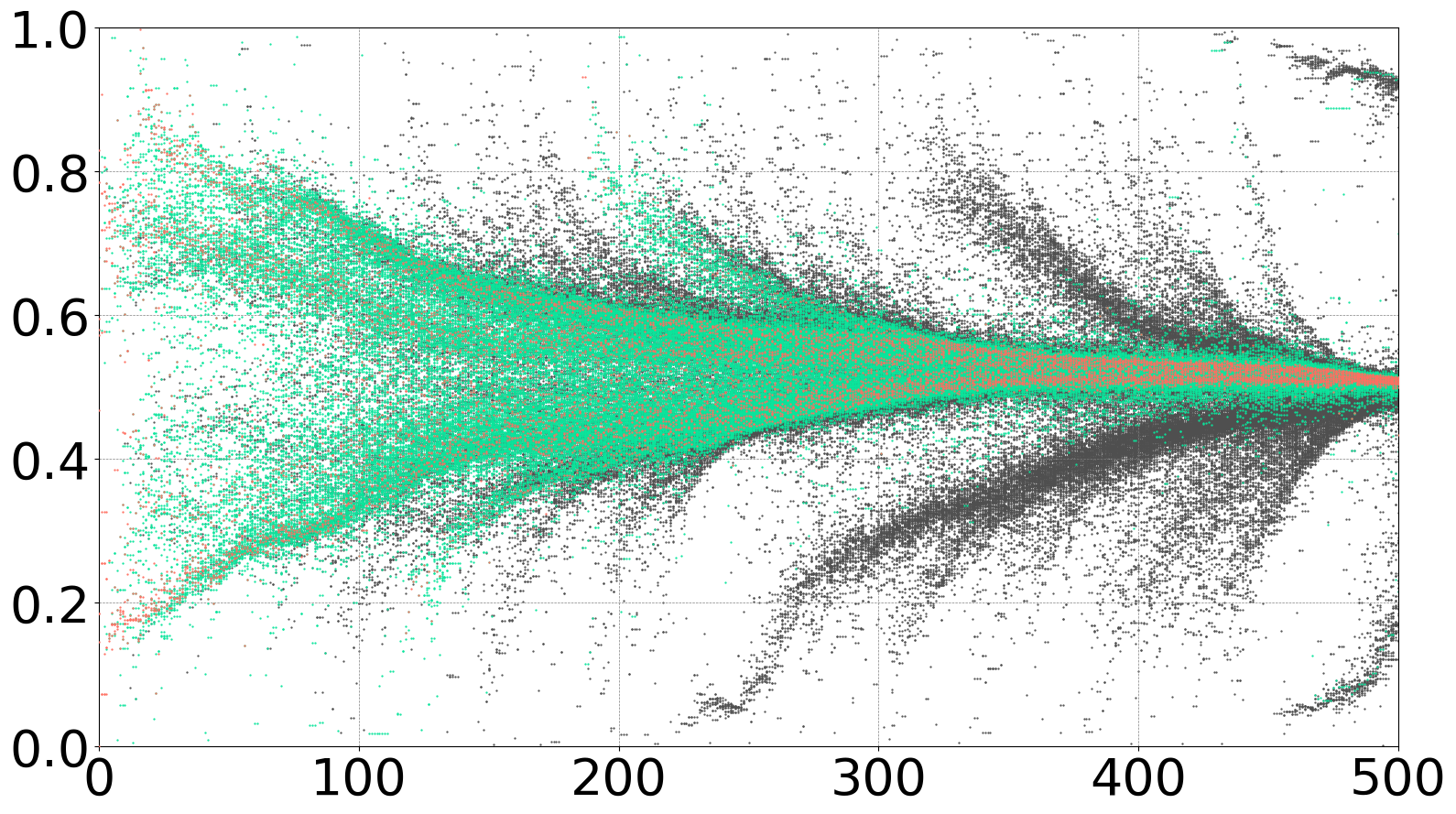}} \\ 
        & Time & Time
    \end{tabular}
    \caption{Representative simulation results in which opinion cascades happen. The confidence bound is $\epsilon = 0.3$, the number of links of new-coming agents is $m =3$, the influence parameter is $\mu = 0.5$, the initial number of nodes in the network is $N_0 = 10$, and the number of agents added to the network per iteration is $\delta = 3$.}
    \label{fig:simulations}
\end{figure}

Simulations running on the aforementioned model show emerging patterns that are significantly different from the ones observed in \cite{GANDICA2024}, which studied the bounded confidence model on growing fully-connected networks. Most notably, in \cite{GANDICA2024}, minor clusters appear and remain stable over time. However, with growing scale-free networks, minor clusters have a different behavior, as shown on Fig. \ref{fig:simulations}. In this case, minor clusters are attracted by major clusters and ultimately merge with them.   

Minor clusters regularly appear over simulations. For $\beta$ (homophily parameter) values in the range $[15,20]$, minor clusters remain relatively stable for a period of time before suddenly merging with the major cluster. However, when $\beta$ is $>20$, the merging process is smoother and slower. This behavior in the merging phenomeon is what we define as "opinion cascades".

The frequence of minor clusters appearing, as well as their size, shape, and merging behavior, are ruled by $\beta$. Note how minor clusters appear only at a distance higher than $\epsilon$ from a main cluster. This happens because nodes outside a cluster's opinion range will remain isolated and eventually form minor clusters over time.

The $\beta$ value range in which both the degree and the opinion similarity determine the link probability ($10 \leq \beta \leq 30$) is our main focus of interest. Our main aim is to understand how, and why, opinion cascades take place.

\subsection{Community structures}

Communities in networks are groups of nodes that are more strongly connected to each other than to the rest of the network. It is possible to think of them as groups of friends in a social network —- people in the same group talk to each other more often than they talk to people outside their group.

The Louvain community detection algorithm \cite{Blondel_2008} defines communities in a network by maximizing modularity, a measure of how strongly connected the nodes are within a community compared to how they would connect at random. It does this by repeatedly merging small communities and refining the structure to maximize modularity.

\begin{figure}[h]
    \centering
     \begin{tabular}{m{0.3cm}cc}
     \rotatebox{90}{\hspace{0.0 cm} Opinion} &\makecell{\includegraphics[width=0.9\linewidth]{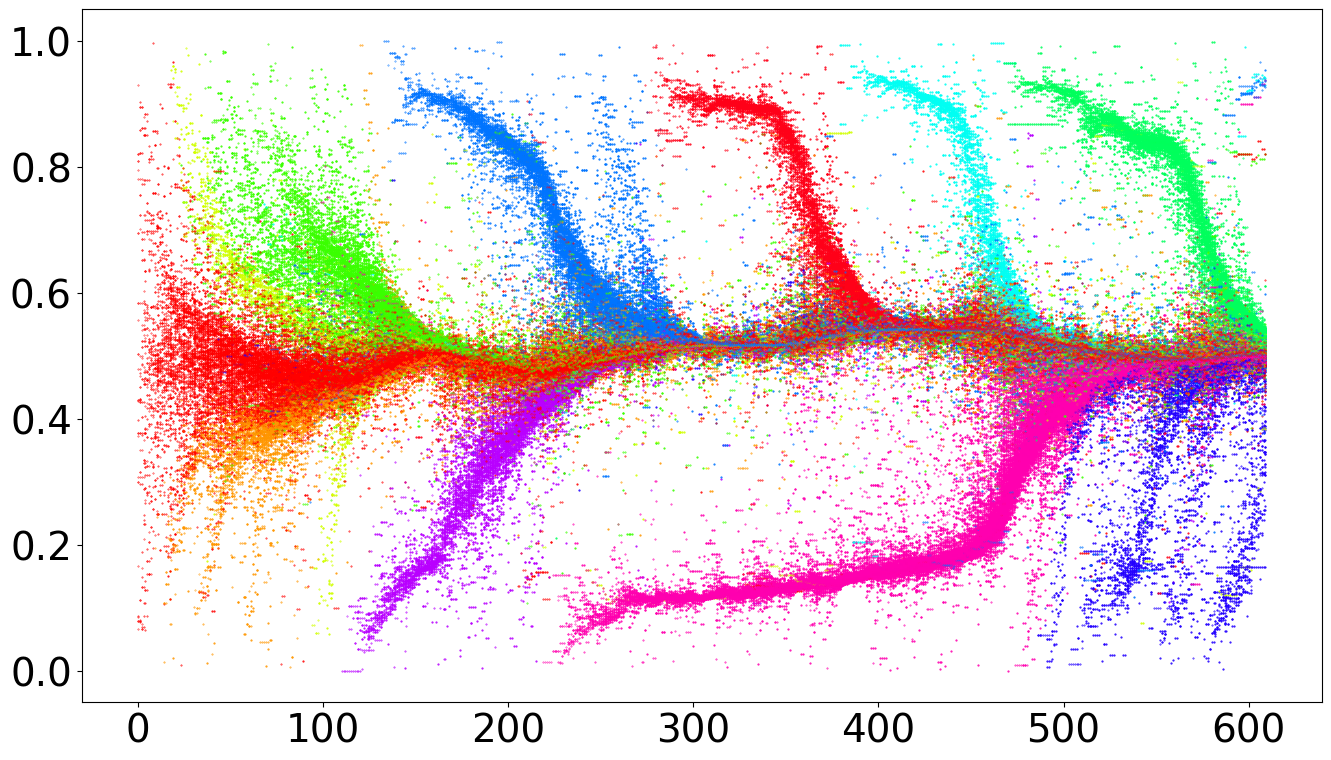}}\\
     & Time
     \end{tabular}
    \caption{Simulation result after detecting communities via the Louvain method. Each community is distinguished by a different colour. The parameters are $\beta=20 , m=3, \epsilon=0.3, \delta=6,$ and $N_0=25$.}
    \label{fig:communities}
\end{figure}

Upon performing the Louvain method after experiments finish, the scale-free networks show well-defined communities, forming one per minor cluster, persisting even after merging with the main cluster. We can see this behavior on Figure \ref{fig:communities}.

\section{Explaining opinion cascades}

In this section, we highlight that opinion cascades happen when an agent added to the network during the simulation is connected to both major and minor clusters. These "Bridge" agents attract the opinions of agents from both clusters, bringing their opinions closer.

In order to showcase this phenomenon and the effect of "Bridge" agents on opinion cascades, we first derive an analytical expression of the probability of a new bridge agent as a function based on its initial opinion. Then, we perform specific simulation experiments assessing the accuracy of the theoretical approximation and explaining the process of opinion cascades.

\subsection{Theoretical probability of agents connected to both minor and major clusters}

We consider the simplified situation involving a major cluster $C_1$ with $n_1$ agents all sharing the same opinion $\theta_1$ and a minor cluster $C_2$ with $n_2$ agents all sharing the same opinion $\theta_2 = \theta_1 + \epsilon + \delta\epsilon$. Remember that $\epsilon$ is the confidence bound  and $\delta\epsilon$ is a positive small value relative to $\epsilon$, ensuring that the minor cluster is outside the main cluster's area of influence. 

Furthermore, we assume that, initially, agents in either cluster are only connected to agents in the same cluster, guaranteeing that there are no links between the two, and therefore no possibility for interaction between them.

Let $P_2(n_1, n_2)$ be the probability that an arriving agent of opinion $x \in [\theta_1, \theta_2]$ is linked to an agent of minor cluster $C_2$, following the definition of the probability of links, we have: 

\begin{align}
    P_2(n_1, n_2) =\frac{\sum_{i \in C_2} k_i \exp(-\beta|\theta_2 - x|)}{\sum_{i \in C_1} k_i \exp(-\beta|\theta_1 - x|) + \sum_{i \in C_2} k_i \exp(-\beta|\theta_2 - x|)}.  
\end{align}

where $k_i$ is the total number of connections to agent $i$. Note that $\sum_{i \in C_1} k_i$ is the total number of oriented connections between the agents of the major cluster $C_1$. The arrival of an agent connected to agents of $C_1$ only creates $2 m$ oriented connections. Of course, the same is true for $C_2$. Therefore, we have:

\begin{align}
    \sum_{i \in C_1} k_i = 2 m n_1, \\
    \sum_{i \in C_2} k_i = 2 m n_2.
\end{align}

Then, for $x \in [\theta_1, \theta_2]$:

\begin{align}
    P_2(n_1, n_2) &=\frac{n_2 \exp(-\beta(\theta_2 - x))}{n_1 \exp(-\beta(x - \theta_1)) + n_2 \exp(-\beta(\theta_2 - x))}  \\
    &= \frac{1}{1 + \frac{n_1}{n_2}\exp(\beta( \theta_1 + \theta_2 - 2x))}.
\end{align}

Replacing $\theta_2$ by its value, we get:
\begin{align}
  P_2(n_1, n_2) = \frac{1}{1 + \frac{n_1}{n_2} \exp(\beta(2\theta_1 + \epsilon + \delta \epsilon - 2x))}  
\end{align}

Moreover, the probability that the arriving agent connects to $C_1$ is:

\begin{align}
  P_1(n_1, n_2) = 1 -  P_2(n_1, n_2).
\end{align}

Then, the probability that a new agent makes m connections only with the minor cluster $C_2$ is:
\begin{align}
    P_{2,m} = P_2(n_1, n_2).P_2(n_1, n_2-1)....P_2(n_1, n_2-m+1).
\end{align}
In this formula, the decreasing from $n_2$ to $n_2 - m+1$ is due to the impossibility to connect twice to the same agent. Therefore, if the new agent is connected to $i$ agents of minor cluster $C_2$ it has only $n_2 - i$ possibilities of connection in $C_2$ left.

Similarly, the probability that a new node makes m connections only with $C_1$ is:

\begin{align}
    P_{1,m} = P_1(n_1, n_2).P_1(n_1-1, n_2)....P_1(n_1-m+1, n_2).
\end{align}

Hence, the probability $P_b$ to get a new node which is connected to both clusters (i.e. is a bridge agent) is:

\begin{align}
    P_b = 1 - P_{1,m} - P_{2,m}.
\end{align}

The next section provides several examples of curves derived from these formulas.

\subsection{Replicating cluster merging from ideal situation and comparing results with theory}

We perform simulation experiments in which we start from the previous simplified setting, and then check how the theoretical probability curves compare to the actual distribution of agents added to the network during simulations. For these experiments, the confidence bound is $\epsilon = 0.3$, the number of links created when agents arrive into the network is $m = 3$, and the number of agents arriving by time step is $\delta = 1$.

\begin{figure}[h!]
    \centering
    \begin{tabular}{m{0.3cm}cc}
    & (a) $\beta = 15$, $t = 0$  & (b) $\beta = 30$, $t = 0$ \\
         \rotatebox{90}{\hspace{0.0 cm} Probability} &\makecell{\includegraphics[width=0.45\linewidth]{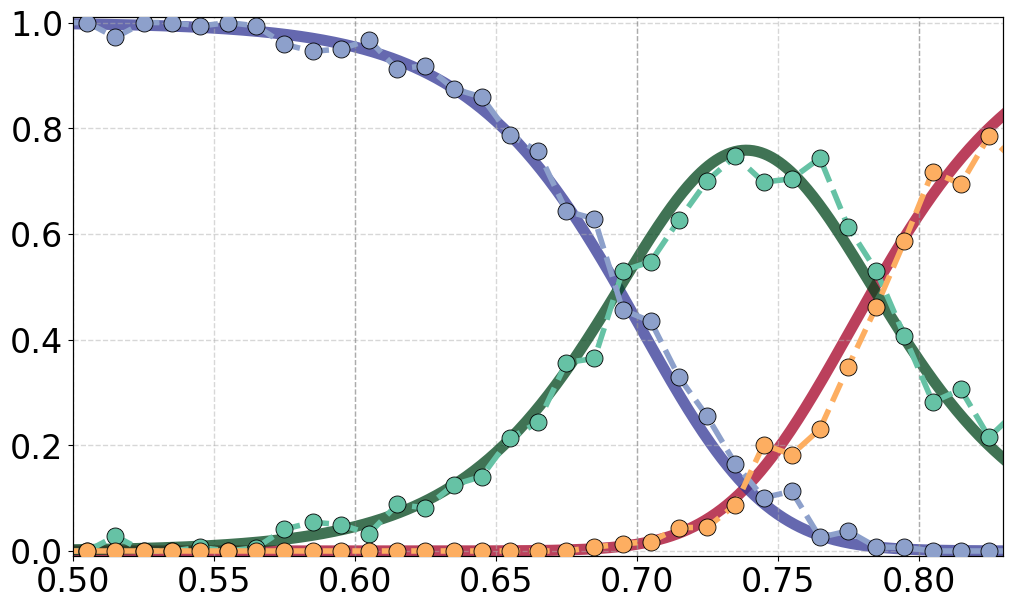}} & 
        \makecell{\includegraphics[width=0.45\linewidth]{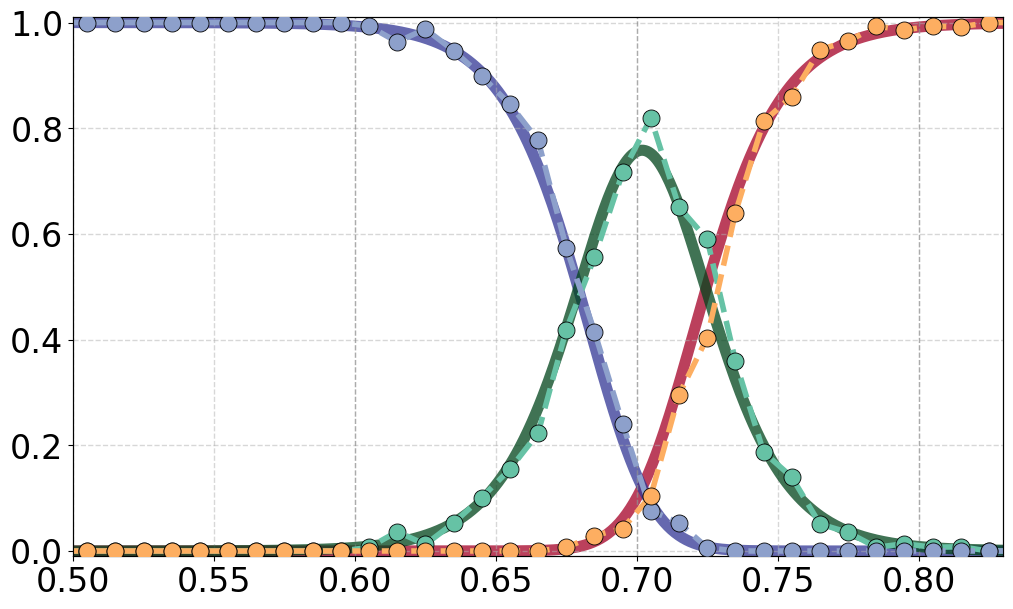}} \\ 
         &&\\
   & (c) $\beta = 15$, $t = 50$  & (d) $\beta = 30$, $t = 50$ \\
         \rotatebox{90}{\hspace{0.0 cm} Probability} &\makecell{\includegraphics[width=0.45\linewidth]{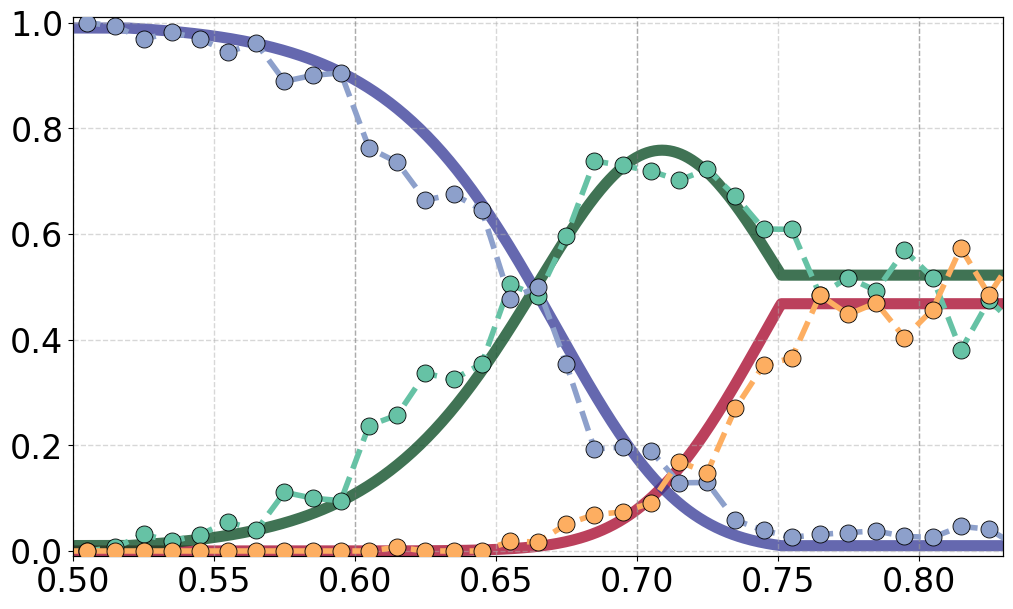}} & 
        \makecell{\includegraphics[width=0.45\linewidth]{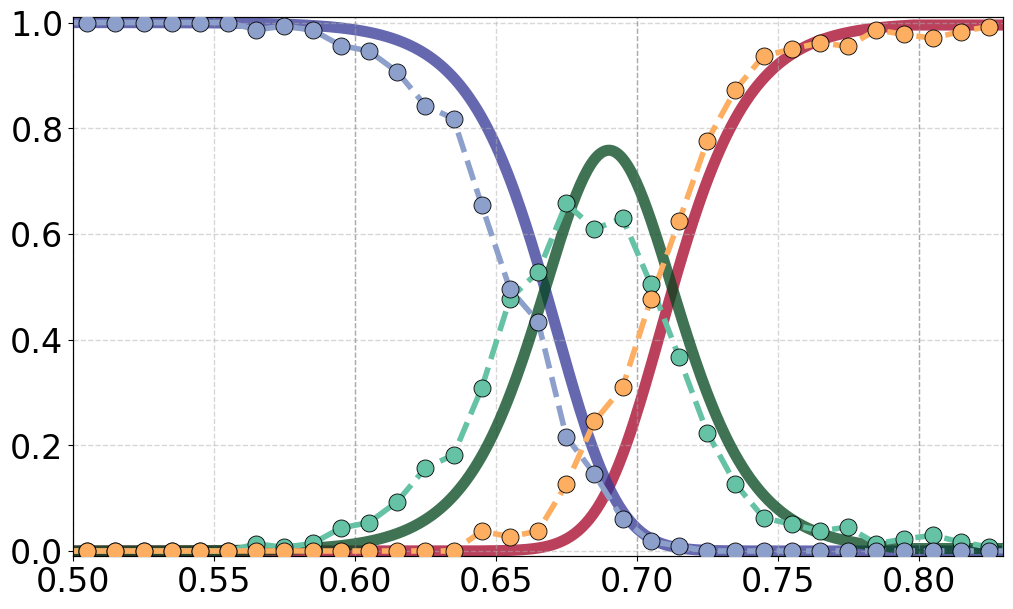}} \\ 
         &&\\
          & (e) $\beta = 15$, $t = 100$ & (f)  $\beta = 30$, $t = 100$\\
         \rotatebox{90}{\hspace{0.0 cm} Probability} & \makecell{\includegraphics[width=0.45\linewidth]{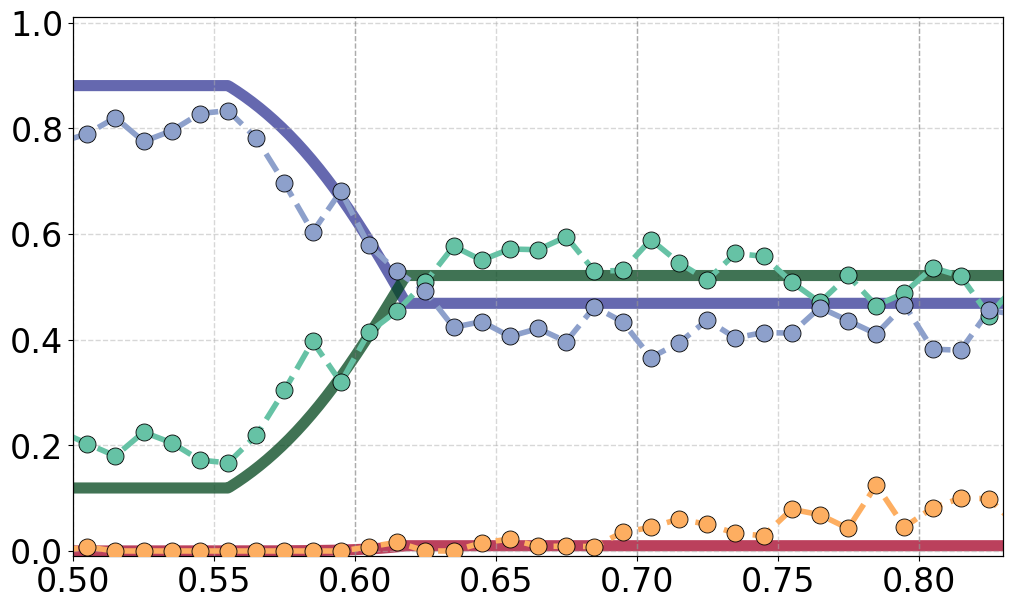}} & 
        \makecell{\includegraphics[width=0.45\linewidth]{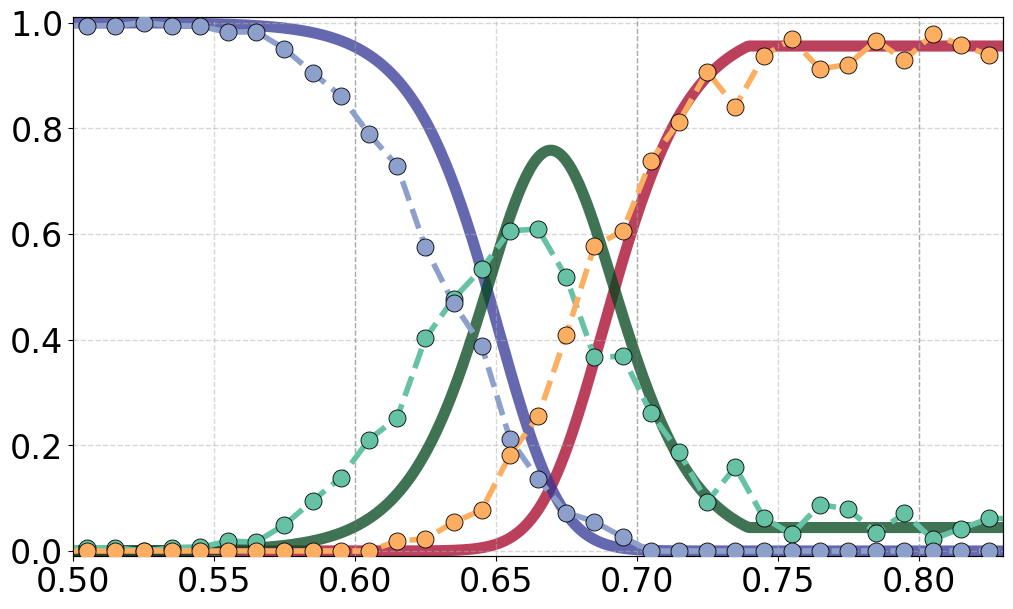}} \\ 
         & Opinion & Opinion\\
    \end{tabular}
    \includegraphics[width= \linewidth]{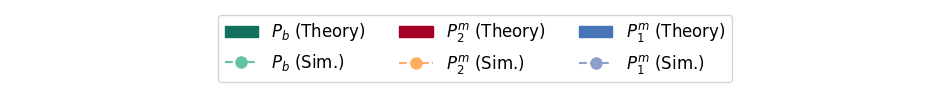}
    \caption{Theoretical probability distributions and experimental frequencies of an arriving agent by type. $P_1^m$: only connected to the major cluster $C_1$, $P_2^m$: only connected to the minor cluster $C_2$, and $P_b$: being a "Bridge". We show results for different times ($t$) and two values of $\beta = 15$ and $30$. Frequencies are averaged over 10,000 replicas and theoretical probabilities are computed with average over the 10,000 replicas of positions and sizes of clusters at time step $t$. 
    }
    \label{fig:probabilities_iterations}
\end{figure}

\begin{figure}[h]
    \centering
    \begin{tabular}{m{0.3cm}cc}
    & (a) $\beta = 15$ & (b) $\beta = 30$\\
         \rotatebox{90}{\hspace{0.0 cm} Number of agents} &\makecell{\includegraphics[width=0.45\linewidth]{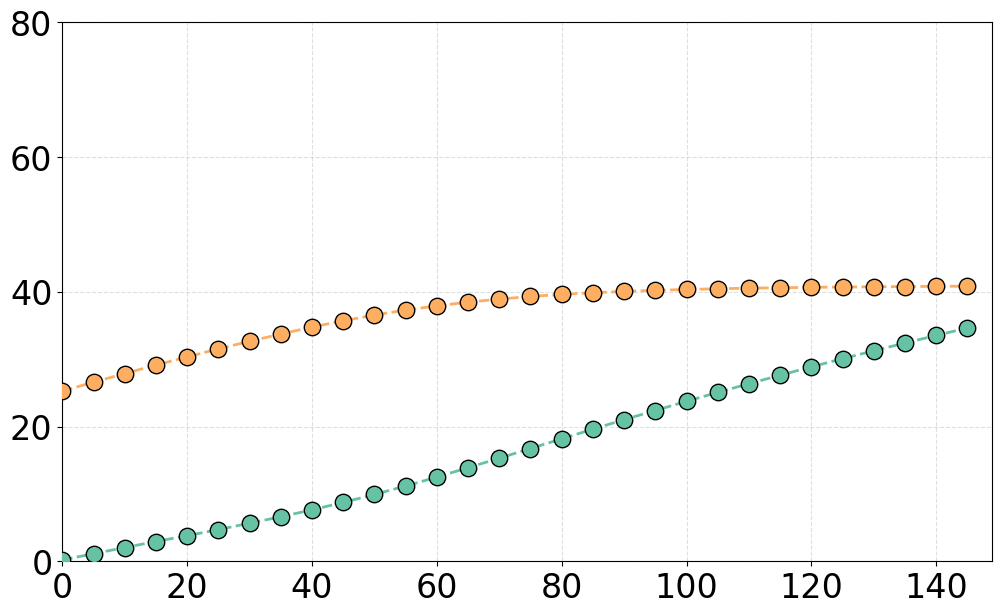}} & 
        \makecell{\includegraphics[width=0.45\linewidth]{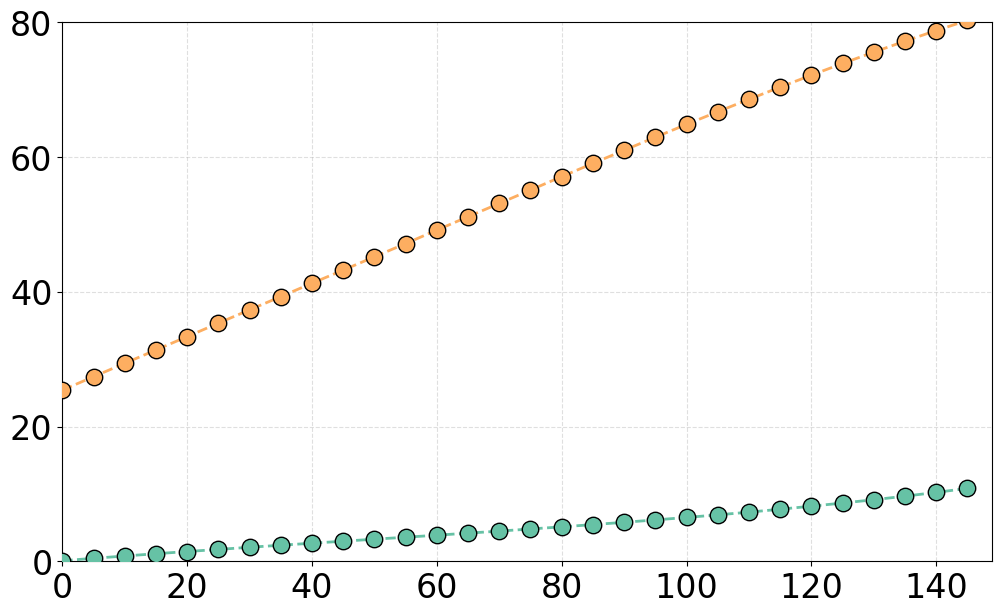}} \\ 
        & Time & Time
    \end{tabular}
    \includegraphics[width=0.5\linewidth]{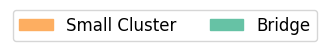}
    \caption{Average (over the 10,000 replicas of experiments) number of "Bridge" agents (green) and agents only connected to the minor cluster $C_2$ (orange), for  $\beta = 15$ and $\beta = 30$.  }
    \label{fig:cluster_sizes}
\end{figure}

\begin{figure}[h]
    \centering
    \begin{tabular}{m{0.3cm}cc}
    & (a) $\beta = 15$ & (b) $\beta = 30$\\
         \rotatebox{90}{\hspace{0.0 cm} Opinion} &\makecell{\includegraphics[width=0.45\linewidth]{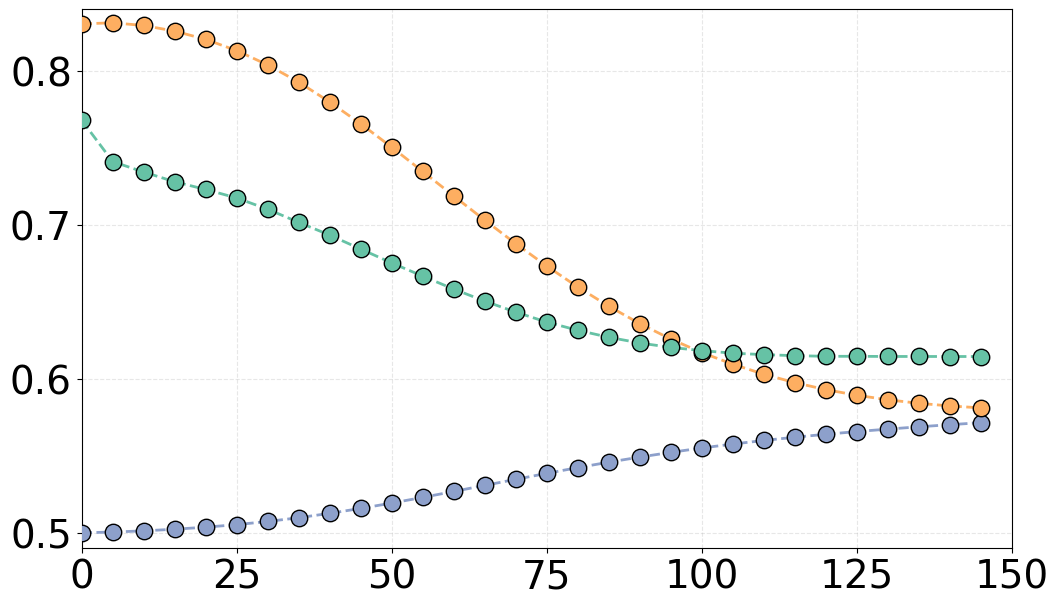}} & 
        \makecell{\includegraphics[width=0.45\linewidth]{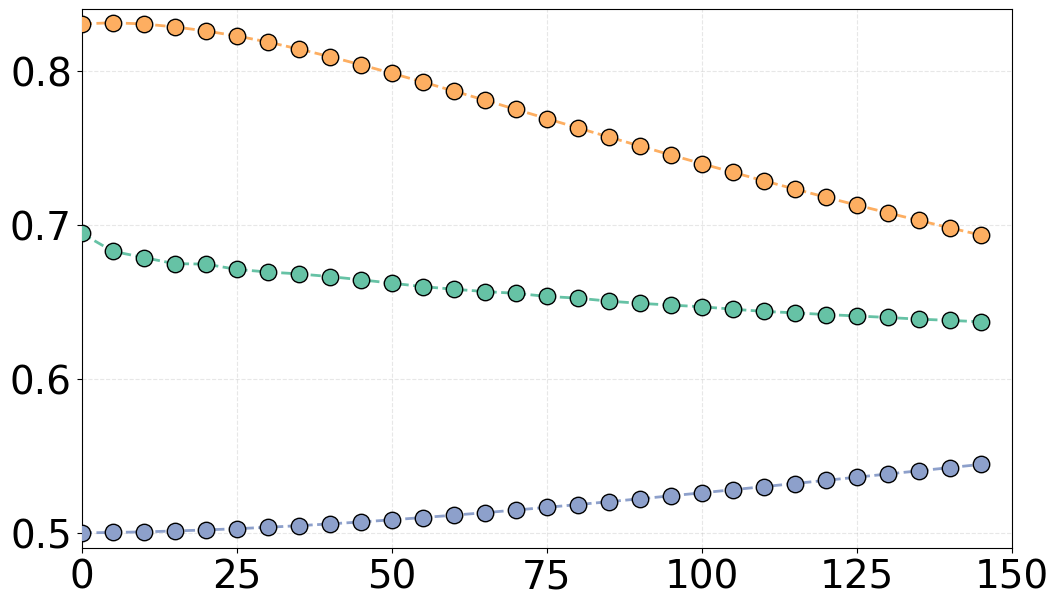}} \\ 
        & Time & Time
    \end{tabular}
    \includegraphics[width=0.7\linewidth]{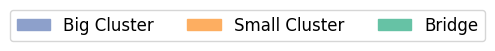}
    \caption{Average (over the 10,000 replicas of experiments) opinions of agents by type (only connected to $C_1$ in blue, only connected to $C_2$ in orange and the Bridges are represented in green), for $\beta = 15$ and $\beta = 30$. }
    \label{fig:cluster_positions}
\end{figure}

The initial setting for these simulations consists of:
\begin{itemize}
    \item Major cluster ($C_1$) with $n_1 = 225$ agents sharing the same opinion $\theta_1 = 0.5$ and connected only to agents of this cluster.
    \item Minor cluster ($C_2$) with $n_2 = 25$ agents sharing the same opinion $\theta_2 = 0.83$ and only connected to agents of this cluster.
\end{itemize}

From this initial state, we run the opinion dynamics algorithm described previously during $T$ time steps ($T = 150$ in this case). We store data about the agents in the network, and start over again from the initial state. We repeat this process $10,000$ times, for $\beta$ values of $15$ and $30$.

For each simulation, we classify the arriving agents based on their links: a) Only connected to $C_1$, b) Only connected to $C_2$, and c) Connected to both ("Bridge"). Then, we divide the opinion space between $\theta_1$ and $\theta_2$ in bins spanning $0.1$ on the opinion space, and place the arriving agents accordingly. Finally, we compute the proportion of each type of agent in each interval over the $10,000$ replicas for each time step.

Figure \ref{fig:probabilities_iterations} shows the frequencies of the different types of arriving agents, obtained over 10,000 replicas of the experiment, for $\beta = 15$ and $\beta = 30$ at 0, 50 and 100 time steps, along with the corresponding theoretical probabilities of the different types of arriving agents, computed with the average positions of the clusters and their average number over the 10,000 replicas at each time step number, in the ideal case where the clusters are disconnected.

The theoretical probabilities fit quite very accurately the experimental values for $t = 0$ and quite well for $t = 50$, but this accuracy decreases significantly at $t = 100$. This could be expected, because the probabilities are computed assuming that all the opinions are the same in each cluster, which is less and less the case while the number of interactions increases. Nevertheless, despite the very strong simplifications for their derivation, the theoretical probabilities express correct qualitative tendencies.

A closer analysis of these tendencies can explain the different patterns of cluster merging shown for $\beta = 15$ and $\beta = 30$ on Figure \ref{fig:simulations}.
Indeed, on Figure \ref{fig:probabilities_iterations}, for $\beta = 15$, from $t = 50$, the probability of arriving bridge agents increases and gets significantly higher than the probability of agents only connected to the small cluster. For $\beta = 30$, this probability of arriving agents only connected to the small cluster remains higher than the probability of arriving bridge agents. 
In particular, at $t = 100$, for $\beta = 15$, the agents arriving with an opinion higher than $0.6$ have a probability of around $0.55$ to be bridge agents and a probability almost zero to be only connected to the small cluster $C_2$. For $\beta = 30$, the probability to of arriving bridge agent has a peak of probability about 0.6, between opinions 0.6 and 0.70, but for opinions higher than 0.75, the probability of arriving bridges is close to 0, while the probability to be only connected to $C_2$ is close to 1. 

Overall, as shown on Figure \ref{fig:cluster_sizes} for $\beta = 15$, the number of bridges increases more rapidly than the number of agents only connected to small cluster $C_2$, especially when the clusters get closer. As a result, the number of bridge agents gets similar to the number of agents only connected to $C_2$. This makes the bridge agents able to attract the opinions of agents only connected to small cluster $C_2$ towards the big cluster $C_1$ very powerfully. Figure \ref{fig:cluster_positions} shows indeed a quick move of the small cluster's average opinion towards the big cluster's average opinion from $t=40$ to $t=100$. 

By contrast, for $\beta = 30$,  Figure \ref{fig:cluster_sizes} shows that the number of bridge agents increases more slowly than the number of agents only connected to small cluster $C_2$. As a result, as shown on Figure \ref{fig:cluster_positions}, the merging process is much slower than for $\beta = 15$.

From this analysis, we can conclude that the cascades are actually due to the rapid increase of bridges that takes place when $\beta < 20$ as soon as the clusters get a bit closer to each other than the confidence bound. This increase is significantly slower for $\beta > 30$ and so is the merging process.

\section{Discussion - conclusion}

Overall, in spite of the simplifications made, the theoretical approximations capture essential qualitative features of the opinion cascading process and allow us to explain the various merging patterns.

For example, when $\beta = 15$, and the minor and major clusters are separated by a distance slightly smaller than the confidence bound $\epsilon$ (i.e. Figure \ref{fig:probabilities_iterations} (a) and (b) ), the probability of arriving bridge agents is significantly higher than the probability of agents only connected to the small cluster. As a result, the number of bridges increases more rapidly and the influence of the bridge agents, attracting the clusters to each other, gets rapidly preponderant. 

By contrast, for $\beta = 30$, in the same situation, the number of arriving agents only connected to $C_2$ remains significantly higher than the number of arriving bridge agents. As a result, the influence of bridge agents remains small (and even decreases over time) and the merging process is much slower than in the previous case.

It is important to underline that this explanation of the cascades requires that the number of connections $m$ is rather small, ensuring a high probability that the bridge agent interacts significantly with agents of both the big and the small clusters. Indeed, if $m$ is large, the bridges are likely to have a much larger number of connections with the big cluster and hence it is likely to interact almost exclusively with agents of that big cluster, thus not affecting the small one. This is the reason why the minor clusters remain stable when agents are fully connected \cite{GANDICA2024}.

To conclude, this work confirms that the interplay between network building and opinion dynamics can generate specific patterns, including opinion cascades. For some values of the parameters, it underlines the difficulty for minor opinion groups to self-sustain in the long run without being absorbed by major groups, when the groups have similar confidence bounds (which can be interpreted as similar openness to others' views). Our work suggests that minor groups should have a significantly smaller confidence bound in order to maintain themselves, which can be interpreted as radicalising. 




\bibliographystyle{IEEEtran}
\bibliography{references}

\end{document}